\documentclass[preprin]{aastex631}
\usepackage{amsmath}
\usepackage{rotating, multirow, ctable,xcolor,isomath}
\usepackage{array,stix2}

\newcommand{\lapprox} {\,
\lower3pt\hbox{$\sim$}\llap{\raise2pt\hbox{$<$}}\,}
\newcommand{\gapprox}{\,\lower3pt\hbox{$\sim$}\llap{\raise2pt\hbox{$>$}},}

\graphicspath{{./}{./images/}} 
 
\shorttitle{Type III Electron Beam Deceleration}
\shortauthors{Azzollini et al.}

\begin{document}

\title{A Multi-Spacecraft Analysis and Modelling of Type III Radio Burst Exciter Deceleration in Inhomogeneous Heliospheric Plasma}

\author[0009-0001-7368-0938]{Francesco Azzollini}
\affiliation{School of Physics \& Astronomy, University of Glasgow, Glasgow, G12 8QQ, UK}

\author[0000-0002-8078-0902]{Eduard P. Kontar}
\affiliation{School of Physics \& Astronomy, University of Glasgow, Glasgow, G12 8QQ, UK}

\begin{abstract}

Electron beams accelerated in solar flares 
and escaping from the Sun along open magnetic field lines 
can trigger intense radio emissions known as type III solar 
radio bursts. 
Utilizing observations by Parker
Solar Probe (PSP), STEREO-A (STA), 
Solar Orbiter (SolO), and Wind spacecrafts, 
the speeds and accelerations of type III exciters are
derived for simple and isolated type III solar bursts. 
For the first time, 
simultaneous four spacecraft observations 
allow to determine positions, 
and correct the resulting velocities 
and accelerations for the location between the
spacecraft and the apparent source. 
We observe velocities and acceleration to change 
as $u(r) \propto r^{-0.37 \pm 0.14}$ 
and $a(r) \propto r^{-1.71 \pm 0.20}$ 
with radial distance from the Sun $r$.
To explain the electron beam deceleration, 
we develop a simple gas-dynamic description 
of the electron beam moving through plasma with
monotonically decreasing density. 
The model predicts that the beam velocity decreases 
as $u(f)\propto f^{1/4}(r)$, so the acceleration
changes $\propto r^{-1.58}$ 
(and speed as $\propto r^{-0.29}$) for the
plasma density profile $n(r)\propto r^{-2.3}$. 
The deceleration is consistent with the average 
observation values corrected for the type III source locations. Intriguingly, the observations also show
differences in velocity and acceleration 
of the same type III observed by different spacecrafts. 
We suggest the difference could be related to
the additional time delay caused by radio-wave scattering between the spacecraft and the source.
\end{abstract}


\section{Introduction}

Solar flares often accelerate electron beams into the solar 
corona that can escape into the interplanetary 
space following magnetic field lines.
These beams interact with the surrounding plasma, 
generating Langmuir
waves, which in turn produce intense radio bursts 
known as type III solar radio bursts
\citep{1950AuSRA...3..541W,1970SoPh...15..222F,1974SSRv...16..189L,1985SoPh..100..537L,2011SSRv..159..107H,2017LRSP...14....2B}.
The emission observed could be either fundamental (F) 
or harmonic (H), depending on whether the emitted frequency 
is close to the local electron plasma frequency 
or double the electron plasma frequency
$f_{\text{pe}}={\omega_{\text{pe}}}/{2 \pi}$, 
where $\omega_{\text{pe}} = \sqrt{4\pi e^2n_e/m_e}$, 
with background electron plasma density $n_e$
and electron mass $m_e$. Early observations of Type III burst 
drift rates suggest that high velocity electron beams with the speed $\sim 0.3c$, where $c$ is the speed of light, 
are required as the exciters to account for the high 
frequency drift rates in the corona
\citep{1950AuSRA...3..541W,1958SvA.....2..653G}. 
The electron beams generating Langmuir waves 
are believed to move through plasma forming a slowly expanding 
beam-plasma structure \citep[][]{2024ApJ...976..233K}.  
While the speeds of the electrons generating 
coronal type III burst is $\sim 30$~keV 
\citep[e.g.][]{1995ApJ...455..347A}, 
the electrons exiting Langmuir waves at 1~au 
are typically a few keV \citep{1985SoPh..100..537L}, 
suggesting type III generating electron deceleration from $\sim c/3$ to well below $\sim 0.1 c$.
However, the exact deceleration radial profile is poorly understood. To address this question, 
extensive work has been made on estimating the velocity 
of type III burst exciters observationally. 
Using two electron density models, one obtained from radio observations and one obtained from the minimum
distance from the Sun permitted by the measured 
arrival direction of the radio signal, \citet{1972Sci...178..743F} find the exciter to decelerate
by a factor of about 2 over distances from $10R_\odot$ 
out to 1~au.
Interestingly, \citet{1987A&A...173..366D} have 
studied 28 type III burst events in the 30-1980 kHz range, associated with detections of Langmuir waves and fast electrons. 
They determine the onset and peak
times for each frequency 
and derive the speeds of electrons exciting
type III bursts finding no significant difference between exciting
electron speeds near and far from the Sun in 12 of the bursts they
studied (43$\%$ of events), 
concluding that there is no strong case in
favor of exciter deceleration. 
\citet{1996AIPC..382...62P} find that the
beam energy decreases by a factor of $\approx 3$ 
from the corona to the interplanetary medium (0.03~au) 
and then remains about constant afterwards. 
More recently, \citet{2015A&A...580A.137K} performed a
statistical survey over 29 simple and isolated IP type III bursts
observed by the STEREO spacecrafts over the 0.1-1~MHz 
frequency range and found that median values 
of the exciter speeds decrease from 0.09c to
0.04c and from 0.16c to 0.09c, with a median deceleration 
of -7 km s$^{-2}$ and -12 km s$^{-2}$, 
for the F and H component, respectively. 

Simulations by \citet{2001SoPh..202..131K,2013SoPh..285..217R} show that the speed of electrons responsible for Langmuir wave generation decreases with distance. The effect is 
attributed to decreasing density. 
Using quasi-linear approach, 
\citet{2023SoPh..298...52L} demonstrate that the
electron beam velocity of $0.38c$ at $5R_\odot$ 
decreases as $r^{-0.5}$ to $0.16c$ at $50R_\odot$. 
To compare with observations, the poorly-known spatial location of the electron source in the solar corona and associated time-delay of the type III source could play an important
role in precise velocity/acceleration determinations.
Therefore, the type III observations with spatial 
localization of the source are preferable.

Simultaneous observations of type III solar radio bursts 
from four spacecrafts spread in the heliospheric angle 
allow better localization of the type III source
\citep{1989A&A...217..237L,1997SoPh..172..307H,2008A&A...489..419B}.
Recently \citet{2021A&A...656A..34M,2023A&A...680A...1C,2025NatSR..1511335C} 
have determined the Heliocentric Earth Ecliptic (HEE) longitude \citep[see][for details]{2006A&A...449..791T} 
of radio burst sources by analyzing the peak flux observed 
at each viewing location of the PSP
\citep{2016SSRv..204....7F}, STEREO-A \citep{2005AdSpR..36.1483K}, Solar
Orbiter (SolO) \citep{2020A&A...642A...1M}, and Wind
\citep{1995SSRv...71..231B} spacecraft. The recorded type III solar
radio burst flux, proportional to $r^{-2}$ was scaled to 1 AU. 
The direction of maximum emission was found by assuming 
that the directivity follows an exponential shape

\begin{equation}\label{eq:musset}
    I_\text{s/c}=I_0 \exp \left(\frac{\cos \left(\theta_{\text{s/c}} - \theta_0 \right)-1}{\Delta \mu}\right)\,,
\end{equation}
where $\theta_{\text{s/c}}$ and  $\theta_0$ are the angles of the
spacecraft and the type III source, $I_\text{s/c}$ is the peak flux measured by the spacecrafts. 
Using flux observations by three or more
spacecrafts, 
$I_\text{s/c}$, the angular source location  longitude
$\theta _0 $, the peak type III burst intensity $I_0$, 
and the angular spread of type III radiation, 
$\Delta \mu$, is determined.

Utilizing the four spacecraft-observations, we account for the
source-spacecraft angular separation to derive exciter velocities and
accelerations (Section~\ref{sec:obs}). These are compared to predictions
from a kinetic model describing the evolution of an electron plasma
structure propagating through ambient plasma with a negative electron
density gradient, offering insights into the dynamics of Type III burst
exciter. Using these unique four spacecraft observations, we deduce type
III burst exciter speeds and accelerations. Section \ref{sec:theory}
provides a simple but extremely useful insight into the electron beam
deceleration due to decreasing density in the inhomogeneous plasma of
the interplanetary space. The comparison of the observations and the
theoretical model provides good agreement.

\section{Multi-spacecraft observations of type III burst
sources}\label{sec:obs}

In our analysis, we use data recorded by the Low Frequency Receiver (LFR)
of the Radio Frequency Spectrometer (RFS) on PSP
(64 logarithmically spaced channels ranging from 10 kHz to 1.7 MHz, with 4-7 s
time resolution), the S/WAVES High Frequency Receiver (HFR) on STEREO-A (frequency
resolution of 25 kHz and a 35-38 s time resolution), the RPW (HFR)
instrument on Solar Orbiter (SolO) (time resolution of $\sim$ 17s 
and 25 kHz spectral resolution) and the Wind/WAVES aboard the Wind spacecraft 
(time resolution of $\sim$ 60s and 4 kHz spectral resolution).
Background levels, calculated using median values over 10 minutes before
the event, are subtracted from the data to improve the signal-to-noise
ratio.  

\begin{figure}
\centering
   \includegraphics[width=0.85\linewidth]{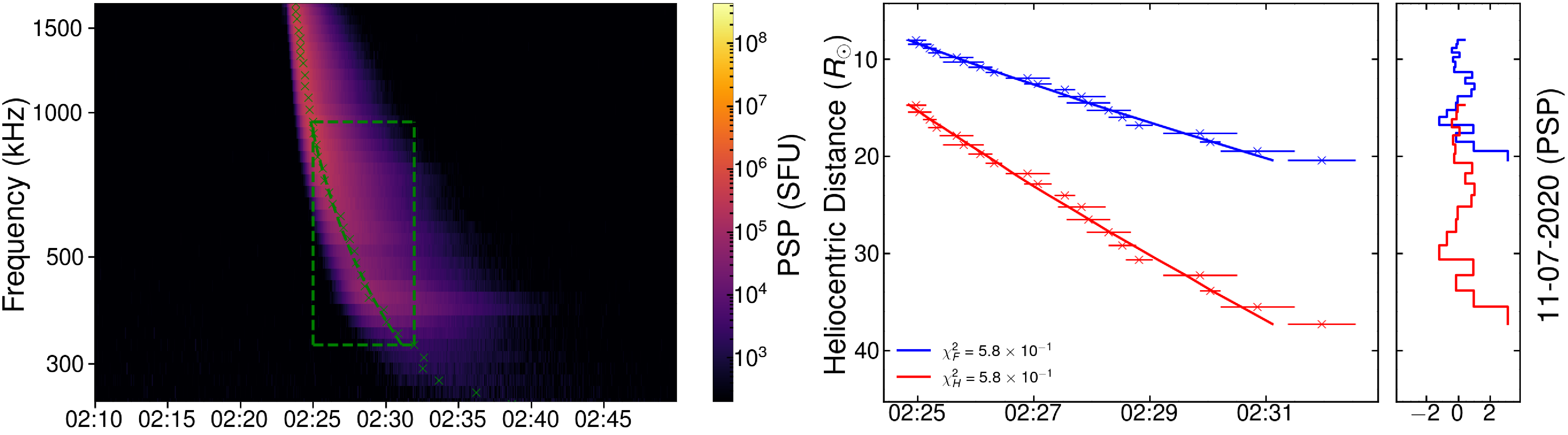}
   \vskip 4pt
   \includegraphics[width=0.85\linewidth]{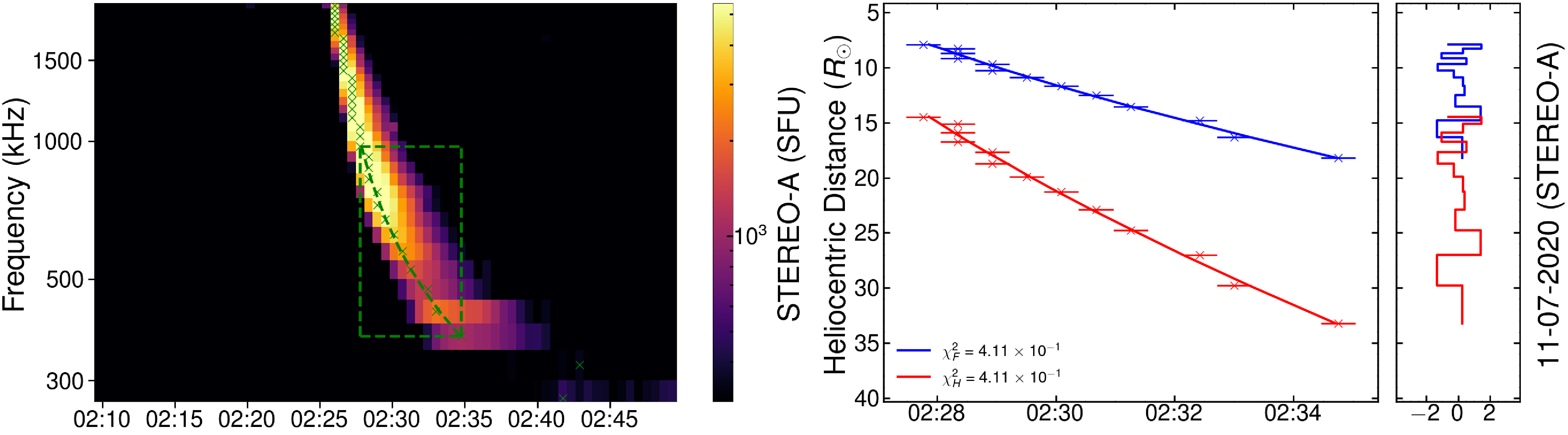}
   \vskip 4pt
   \includegraphics[width=0.85\linewidth]{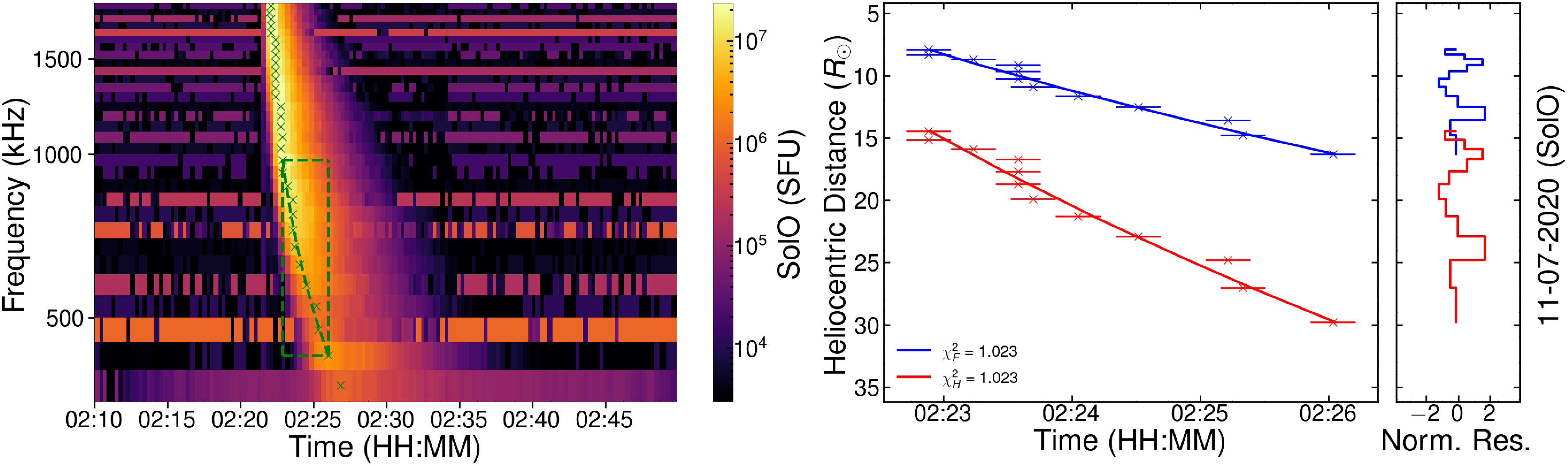}
   \caption{\label{fig:11_07_dynamic_spectra}Dynamic spectra (left) and
  frequency-time (right) on the 11 July 2020 by the PSP, STEREO-A and
  SolO spacecraft (from top to bottom). For each spacecraft, the
  peak-flux frequencies (and the fit) are plotted on the right for the
  times-frequencies selected by the green dashed box, containing peak
  flux points (green 'X' symbols),  along with their fitted curve (green
  dashed line), while the fitted positions of the emitter as a function
  of time and the normalized residuals from the fit are shown on the
  right. Blue and red lines correspond, respectively, to the fundamental and harmonic components.}
\end{figure}

This work investigates type III radio burst events 
recorded on 11 July 2020 (Figure~\ref{fig:11_07_dynamic_spectra}) observed by STEREO-A, PSP and
SolO; frequency drifts are analyzed for frequencies below 1 MHz due to
instrument time resolution. 
The type III burst data  have sufficiently high frequency 
and time resolution, with data points
forming a smooth, monotonic relationship between peak time and frequency
(see right panel in Figure~\ref{fig:11_07_dynamic_spectra}).

\begin{figure}
    \centering
    \includegraphics[width=0.85\linewidth]{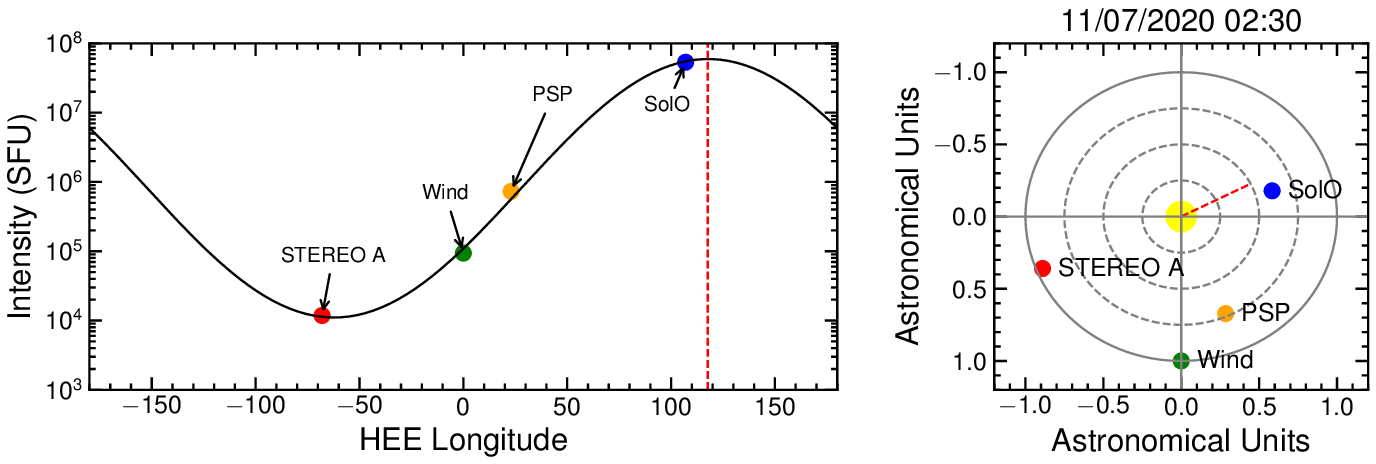} 
    \caption{Type III burst peak fluxes measured by four different
    spacecrafts (left) and spacecraft positions (right) in HEE
    coordinates during the 11 July 2020 (2:30 UT) event. The direction
    of maximum directivity is found by fitting Equation~\ref{eq:musset}
    for the peak fluxes from STEREO-A, PSP, Wind and SolO at 979 kHz.
    This frequency was selected on the assumption that for $\gapprox 1$
    MHz, the Sun's magnetic field is approximately radial, meaning that
    the observed radio sources will mainly have been shifted radially
    due to scattering. On the left peak fluxes are plotted as a function
    of HEE Longitude. The red dashed line shows the position of the
    radio source as revealed by the directivity fit. On the right are
    the position of Solar Orbiter, Parker Solar Probe, STEREO-A and Wind
    projected in the plane of the HEE coordinate system.}
    \label{fig:sc_position_11July}
\end{figure}

% \begin{figure}
%     \centering
%     \includegraphics[width=0.6\linewidth]{Images/stereo_hfr_spec_2020-07-11_02_02.eps}\\%
%     \includegraphics[width=0.6\linewidth]{Images/solo_hfr_spec_2020-07-11_02_02.eps}\\%
%     \includegraphics[width=0.6
%     \linewidth]{Images/wind_waves_spec_2020-07-11_02_02.eps}\\%
%     \includegraphics[width=0.6\linewidth]{Images/psp_hfr_dynspec_2020-07-11_02_02.eps}\\%
%     \caption{Dynamic spectra of the 11 July 2020 02:30 UT type III radio burst observed from STEREO-A, SolO, Wind and PSP.}
%     \label{fig:sc_ds_July21}
% \end{figure}

Observations of the type III burst from a for different spacecrafts
allows us to determine the peak of type III burst directivity 
and hence the source angular location.  Following the approach by
\citet{2023A&A...680A...1C,2021A&A...656A..34M,2025NatSR..1511335C}, 
the source-spacecraft angular separations $\phi = \theta _0 -\theta_{s/c}$ 
are found to be
$\sim$94$^\circ$, $\sim$185$^\circ$ and $\sim$10$^\circ$ for PSP,
STEREO-A and SolO respectively (Figure~\ref{fig:sc_position_11July});
For the range of frequencies between $2-0.5$~MHz considered, the radial
motion of electrons is a good approximation since the curvature of the
Parker spiral at these distances (order of $10R_\odot$, where $R_\odot$
is the solar radius) is rather small.  

\begin{figure}
    \centering
    \includegraphics[width=0.65\linewidth]{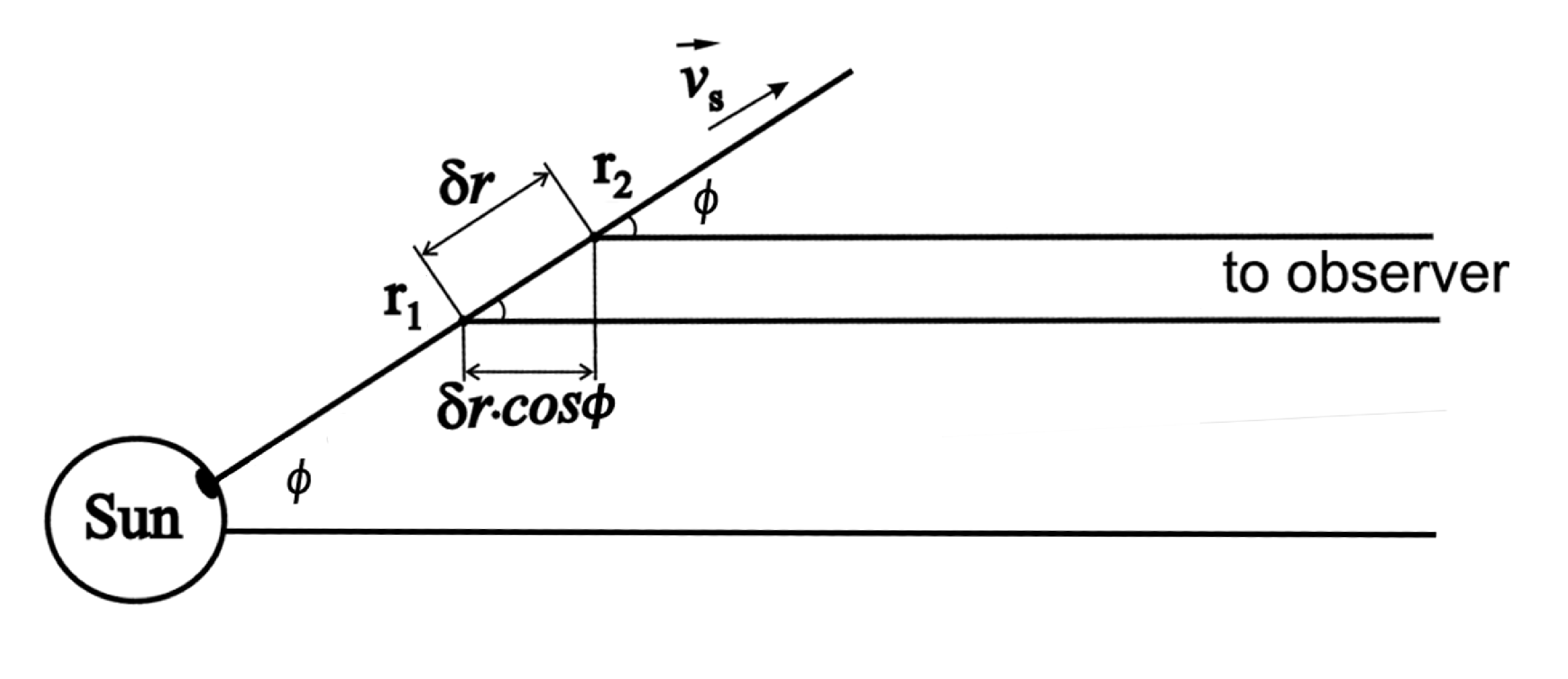}
    \caption{Type III exciter propagating from position ${r_1}$ to
    ${r_2}$ with constant velocity $v_s$ at an angle $\phi$ to the line
    of sight. This simple representation allows us to correct for the
    source-to-spacecraft light travel time (Equation~\ref{eq:delta_t}).}
    \label{fig:melnik}
\end{figure}

To determine the drift rate of type III burst source, one needs to take
into account the direction of exciter motion with respect to the
observing spacecraft, e.g. the angular separation $\phi = \theta _0
-\theta_{s/c}$ that affects the drift rate derivation. Similarly to
\citet{1963ApJ...138..239H,2000SoPh..197..387L,2011SoPh..269..335M,
2015SoPh..290..193M}, if a source is moving radially, with a constant
velocity $v_{\mathrm{s}}$ at an angle $\phi$ to the line-of-sight 
and generating radio-waves at points $r_1$ and $r_2$
(Figure~\ref{fig:melnik}), the time difference between the arrivals of
these waves to the observer is
\begin{equation}\label{eq:delta_t}
    \delta t \approx \frac{\delta r}{v_{\mathrm{s}}}\frac{c-v_{\mathrm{s}} \cos \phi}{c}\,,
\end{equation}
where $\delta r=r_2-r_1$ is the radial distance traveled. Only for
nearly perpendicular to the line-of-sight motion $\delta t \approx
\frac{\delta r}{v_{\mathrm{s}}}$ is unaffected by the radio-wave
propagation. 

Solving Equation (\ref{eq:delta_t}), the speed over the distance $\delta
r$ can be written
\begin{equation}
    \frac{\delta r}{\delta t}=v_s \frac{c}{c-v_s\cos\phi}\,.
\end{equation}
Hence, the drift rate can be written
\begin{equation}\label{eq:df_dt}
\frac{\delta f}{\delta t} = \frac{d f}{d r} \frac{\delta r}{\delta t} 
=\frac{df }{dr} \frac{c v_s}{c-v_s\cos\phi}\simeq \frac{df}{dt}(1+v_s/c \cos\phi),\,
\end{equation}
where $f$ could be either plasma frequency or its harmonic. The second
term in the Equation~\ref{eq:df_dt} presents the correction due to the
radio-wave travel time with speed $c$. One can see that the correction
is larger for larger exciter speeds $v_s$. The correction is also zero
for $\phi =90 ^0$, i.e. the emission travels the same distance and
there is no frequency dependent delay. Note that
\citet{2015A&A...580A.137K} used different angle definition in their
appendix, so that their correction is zero for the deviation angle
$\Delta \phi$.

Following previous research works \citep[e. g.][]{2015A&A...580A.137K},
the drift-rate of type III bursts is determined using the flux maximum
for each frequency (see Figure~\ref{fig:11_07_dynamic_spectra}), i.e.
fitting the frequency as a function of time using the power-law model
\begin{equation} \label{eq:f_fit}
    f_i = A_i \left(t_i-t_{0i}\right)^{B_i} \, [\text{MHz}],
\end{equation} 
where $i = \mbox{F, H}$ depending on whether fundamental or harmonic
emission is considered. Then, the observed frequency can be related to
the spatial location using density model 
\begin{equation}\label{eq:density_r}
    n(r)=1.4 \times 10^6 \left(R_\odot/r\right)^{2.3}\;\mbox{[cm$^{-3}$],}
\end{equation}
which is a power-law fit \citep[see][ for details]{2023ApJ...956..112K}
to the \citet{1960ApJ...132..821P} model with constant temperature and
constants chosen to agree with in-situ density measurements at 1~au
adapted by \citet{1999A&A...348..614M}. Frequency as a function of time
is also a power-law function  
\begin{equation}\label{eq:f_i}
   f_i = C \left(\frac{r_i}{R_\odot}\right)^{D} \simeq 10.53 \left(\frac{r_i}{R_\odot}\right)^{D} \mbox{[MHz]}\,,
\end{equation}
where $C=8.9 \,\sqrt{1.4}$ and $D = -2.3/2$ can be found using density
model. This approximation yields densities within 20\% of the density
models for the range of frequencies considered here (see Figure 11 by
\citet{2023ApJ...956..112K} for the comparison with different density
models). 

Given the density model (Equation \ref{eq:density_r}), one can find the
parameters $\alpha_i$ and $\beta_i$ for the power-law model from
\citet{2015A&A...580A.137K}
\begin{equation}\label{eq:v_fit}
    v_i = \alpha_i \left(\frac{r_i}{R_{\odot}}\right)^{\beta_i} \, ,
\end{equation} 
with $\alpha_i = \frac{B_i}{D}
\left(\frac{A_i}{C}\right)^{\frac{1}{B_i}}$ and $\beta_i = 1 -
\frac{D}{B_i}$ ; as well as parameters $\gamma_i$ and $\delta_i$ from
the exciter acceleration power-law model
\begin{equation} \label{eq:a_fit}
    a_i = \gamma_i \left(\frac{r_i}{R_{\odot}}\right)^{\delta_i} \, ,
\end{equation} 
with $\gamma_i =
\frac{B_i}{D}\left(\frac{B_i}{D}-1\right)\left(\frac{A_i}{C}\right)^{\frac{2}{B_i}}$
and $\delta_i = 1 - \frac{2D}{B_i}$.

\section{Exciter velocity and acceleration}

\begin{figure}
\centering
   \includegraphics[width=0.9\linewidth]{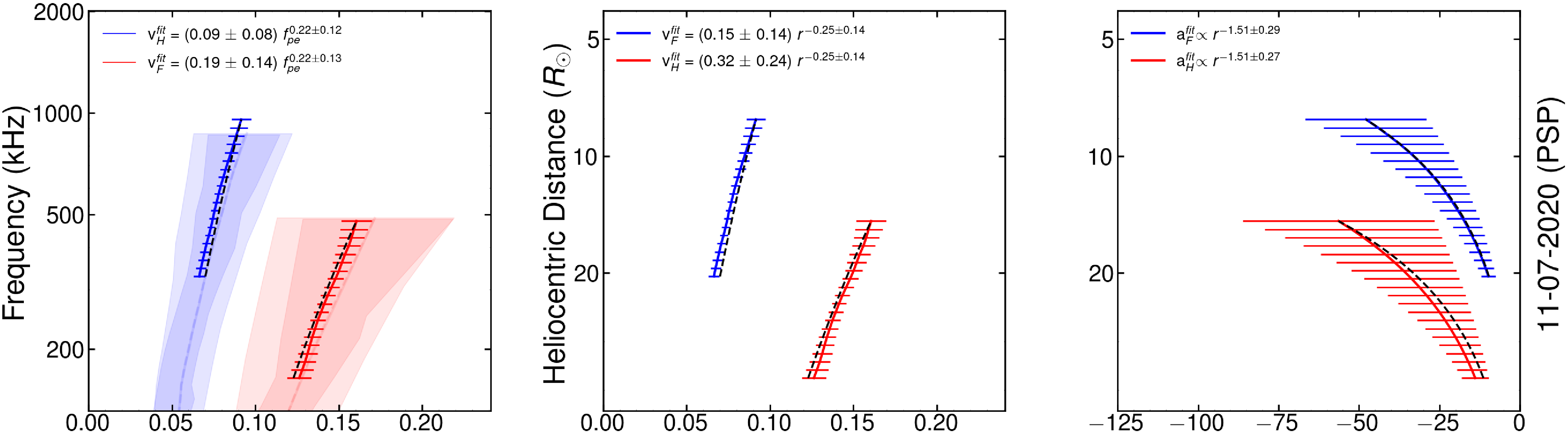}
   \vskip 4pt
   \includegraphics[width=0.9\linewidth]{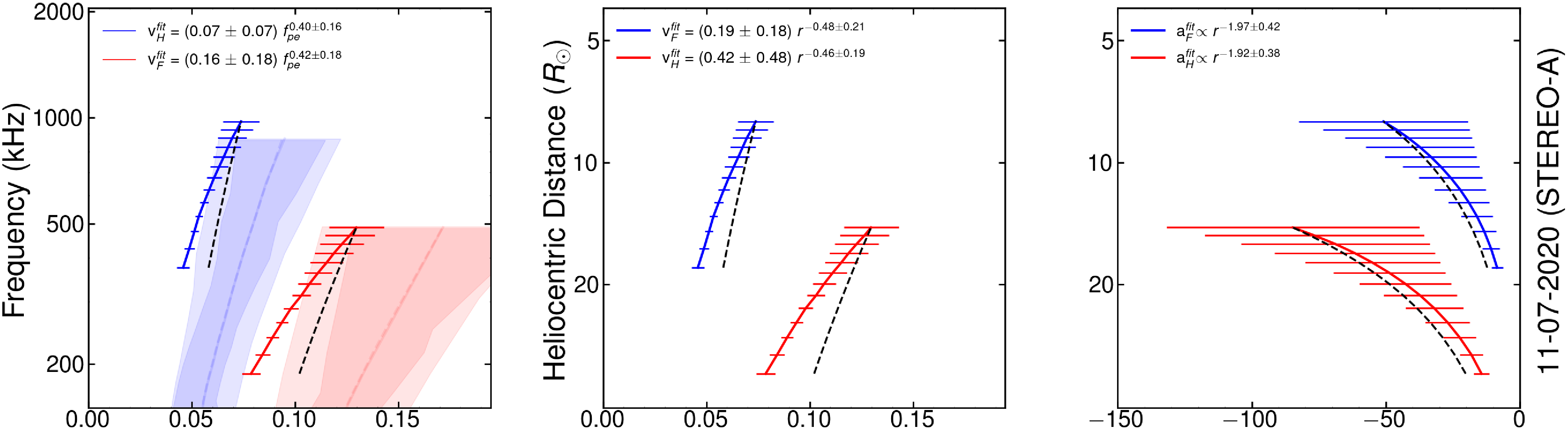}
   \vskip 4pt
   \includegraphics[width=0.9\linewidth]{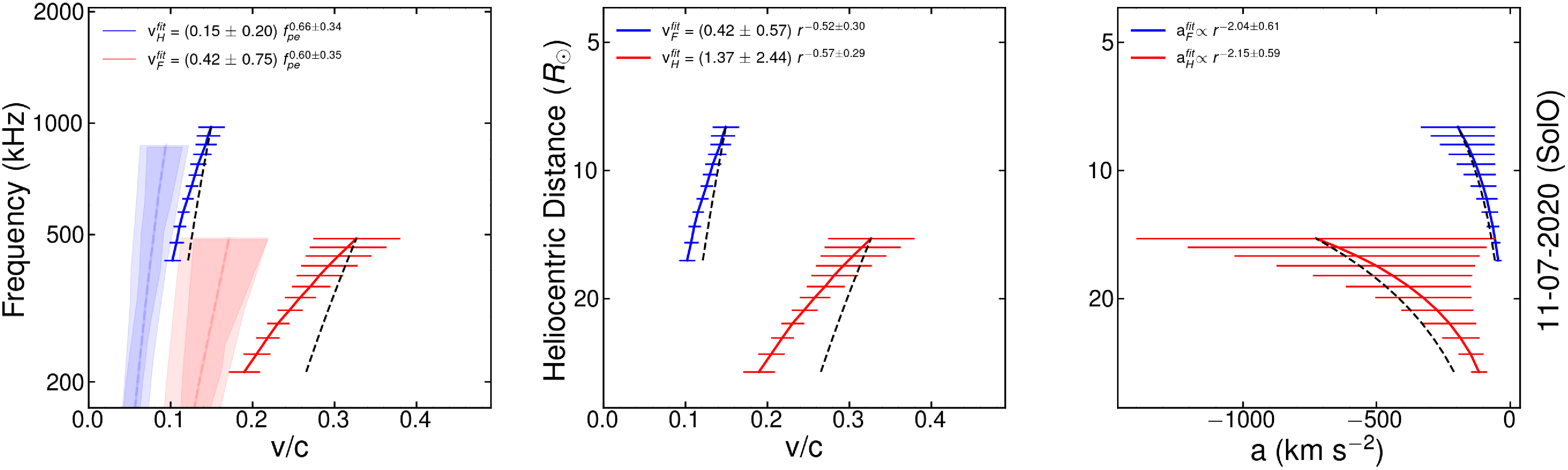}
   \caption{\label{fig:11_07_speed_accel} 
   Exciter velocities and accelerations from the 11 July 2020 type III
   burst for the PSP, STEREO-A and SolO spacecraft (from top to bottom). Blue and
   red lines correspond, respectively, to the fundamental and harmonic components.
   Velocity as a function of frequency is shown on the left, where the shaded areas show velocity deduced from  \citet{2015A&A...580A.137K} by STEREO-A and STEREO-B data. Median values are shown with transparent solid (STEREO-A) and dashed (STEREO-B) lines. On the right, velocity and acceleration
   of the exciter are plotted as a function of distance. 
   Black dashed lines correspond to the result from
   Equation~\ref{eq:u_x}, where $x_0$ corresponds 
   to the location where the highest analysed frequency is emitted.}
\end{figure}

Using the assumptions the equations for velocity and acceleration, one
can fit the peak-flux-frequency versus time with a power-law velocity
and power-law acceleration fits. The main results, where we took into
account source location, are presented in
Figures~\ref{fig:11_07_speed_accel}, while the fit parameters are
presented in Table~\ref{table:params}. 
The additional event observed on July 21, 2020, is presented in Appendix~\ref{app:July21}.

% \begin{figure}
% \centering
%    \includegraphics[width=0.9\linewidth]{Images/stereo_hfr_dynspec_2020-07-21_02_03.png}

%    \includegraphics[width=0.9\linewidth]{Images/stereo_hfr_vel_2020-07-21_02_03.png}
%    \caption{\label{fig:21_07} Data recorded on the 21 July 2020 by the
%    STEREO-A spacecraft. \textbf{Top:} The fitted Type III spectrum is
%    plotted on the left highlighted by the green dashed box, containing
%    peak flux points (green 'X' symbols), along with their fitted curve
%    (green dashed line). The fit for the position of the emitter as a
%    function of time and the normalized residuals from the fit are shown
%    on the right. \textbf{Bottom:} Exciter velocity as a function of
%    frequency is shown on the left, where dotted lines show median
%    velocity values found by \citet{2015A&A...580A.137K}. Velocity and
%    acceleration of the exciter as a function of distance are plotted on
%    the right, with blue and red lines corresponding, respectively, to
%    the fundamental and harmonic components. Black dashed lines
%    correspond to the result of Equation~\ref{eq:u_x}.}
% \end{figure}

Error due to frequency resolution and choice of density model is assumed
to be negligible, while the main source of error was deemed to be the
temporal resolution of the instrument. 
In the case of PSP data, whenever
half the width of the light curve at 90-95\% of the peak exceeded
instrumental time resolution, the former was taken as instrumental
error, allowing to take the uncertainty on radio flux peak time into
account. Uncertainties in derived parameters were estimated using a Markov Chain Monte Carlo (MCMC) method \citep{1986nras.book.....P}. The MCMC parameter samples 
are used to calculate standard deviations 
in velocity and acceleration.

The estimated velocities range from 0.04c at 375 kHz to 0.14c at 1 MHz,
or from 0.10c at 185 kHz to 0.45c at 500 kHz, depending on whether
fundamental or harmonic emission is considered, consistent with
estimates from previous studies. Similarly, median values for $\beta_i$
are found to be $\beta_F \sim \beta_H \sim$ -0.37 $\pm$ 0.15, in
agreement with $\beta_f \sim \beta_H \sim $ -0.35 from the power law
model in \citet{2015A&A...580A.137K}.

Uncertainty in accelerations range between 20-70\%, with exciter accelerations varying from -4 km s$^{-2}$
at 375 kHz to -194 km s$^{-2}$ at 1 MHz for fundamental emission.
Accelerations for harmonic emission are up 4 times greater in magnitude, ranging from
-6 km s$^{-2}$ at 185 kHz up to a minimum of -725 km s$^{-2}$ at
500 kHz. These estimates up an order of magnitude greater than values
found by \citet{2015A&A...580A.137K}. Exciter accelerations are observed
to decrease rapidly, with average values $\delta_F \approx \delta_H
\approx -1.71 \pm 0.20$. Throughout this work, emission is assumed to be either at the fundamental or harmonic frequency. However, as noted by \citet{1987A&A...173..366D}, harmonic emission may begin near the peak of the burst and subsequently become increasingly dominant. Nonetheless, the median $\beta_i$ values derived in this study are expected to remain unaffected.

\begin{deluxetable}{r|cccccc}\label{table:params}
\tablecolumns{5}
\tablewidth{1.0\columnwidth} 
\tablecaption{Parameters $A_i$, $B_i$ and $t_{0i}$ from
Equation~\ref{eq:f_fit}, where $i=F,\,H$ depending on whether
fundamental or harmonic emission is assumed.} \tablehead{Date & $A_F$ &
$B_F$ & $t_{0F}$ & $A_H$ & $B_H$ & $t_{0H}$}
        \startdata
        \multicolumn{7}{c}{PSP} \\
        \hline
        11/07/2020 & 93.3$\pm$59.2 & -0.92$\pm$0.092 & 720$\pm$31 & 54.4$\pm$54.7 & -0.92$\pm$0.12 & 719$\pm$36\\
        \hline
        \multicolumn{7}{c}{STEREO-A} \\
        \hline
        11/07/2020 & 54.4$\pm$54.7 & -0.77$\pm$0.10 & 910$\pm$49 & 30.2$\pm$20.7 & -0.78$\pm$0.09 & 897$\pm$47\\
        \hline
        21/07/2020 & 133.3$\pm$60.05 & -0.94$\pm$0.08 & 1463$\pm$34 & 75.7$\pm$41.1 & -0.95$\pm$0.08 & 1451$\pm$40\\
        \hline
        \multicolumn{7}{c}{SolO} \\
        \hline
        11/07/2020 & 27.8$\pm$34.7 & -0.76$\pm$0.14 & 680$\pm$28 & 11.2$\pm$17.7 & 0.74$\pm$0.13 & 692$\pm$28\\
        \hline
        \enddata
\end{deluxetable}

\begin{deluxetable}{r|cccc}
\tablecolumns{5}
\tablewidth{1.0\columnwidth} 
\tablecaption{Parameters $\alpha_i$ and $\beta_i$ (top) from
Equation~\ref{eq:v_fit}, and $\gamma_i$ and $\delta_i$ (bottom) from
Equation~\ref{eq:a_fit}, where $i=F,\,H$ depending on whether
fundamental or harmonic emission is assumed.} 
\tablehead{ Date & $\alpha_F$ & $\beta_F$ & $\alpha_H$ & $\beta_H$ }
\startdata
        \multicolumn{5}{c}{PSP} \\
        \hline
        11/07/2020 & 0.15$\pm$0.11 & -0.25$\pm$0.13 & 0.32$\pm$0.32 & -0.25$\pm$0.15\\
        \hline
        \multicolumn{5}{c}{STEREO-A} \\
        \hline
        11/07/2020 & 0.19$\pm$0.23 & -0.48$\pm$0.21 & 0.42$\pm$0.46 & -0.46$\pm$0.19 \\
        \hline
        21/07/2020 & 0.13$\pm$0.05 & -0.22$\pm$0.10 & 0.25$\pm$0.12 & -0.21$\pm$0.11 \\
        \hline
        \multicolumn{5}{c}{SolO} \\
        \hline
        11/07/2020 & 0.42$\pm$0.64 & -0.52$\pm$0.27 & 1.37$\pm$2.71 & -0.57$\pm$0.33\\ 
        \hline \hline
        Date & $\gamma_F / 10^4$ & $\delta_F$ & $\gamma_H / 10^4$ & $\delta_H$\\
        \hline
        \multicolumn{5}{c}{PSP} \\
        \hline
        11/07/2020 & -0.07$\pm$0.13 & -1.51$\pm$0.26 & -0.32$\pm$0.78 & -1.51$\pm$0.31\\
        \hline
        \multicolumn{5}{c}{STEREO-A} \\
        \hline
        11/07/2020 & -0.22$\pm$0.58 & -1.97$\pm$0.42 & -1.02$\pm$2.51 & -1.92$\pm$0.38\\
        \hline
        21/07/2020 & -0.05$\pm$0.05 & -1.44$\pm$0.20 & -0.17$\pm$0.22 & -1.42$\pm$0.22\\
        \hline
        \multicolumn{5}{c}{SolO} \\
        \hline
        11/07/2020 & -1.19$\pm$3.90 & -2.04$\pm$0.55 & -13.94$\pm$59.01 & -2.15$\pm$0.67\\
        \hline
\enddata  
\label{tab:schecParams}
%\tablecomments{Table comments.}
\end{deluxetable}

\begin{figure}
\centering

\includegraphics[width=0.9\linewidth]{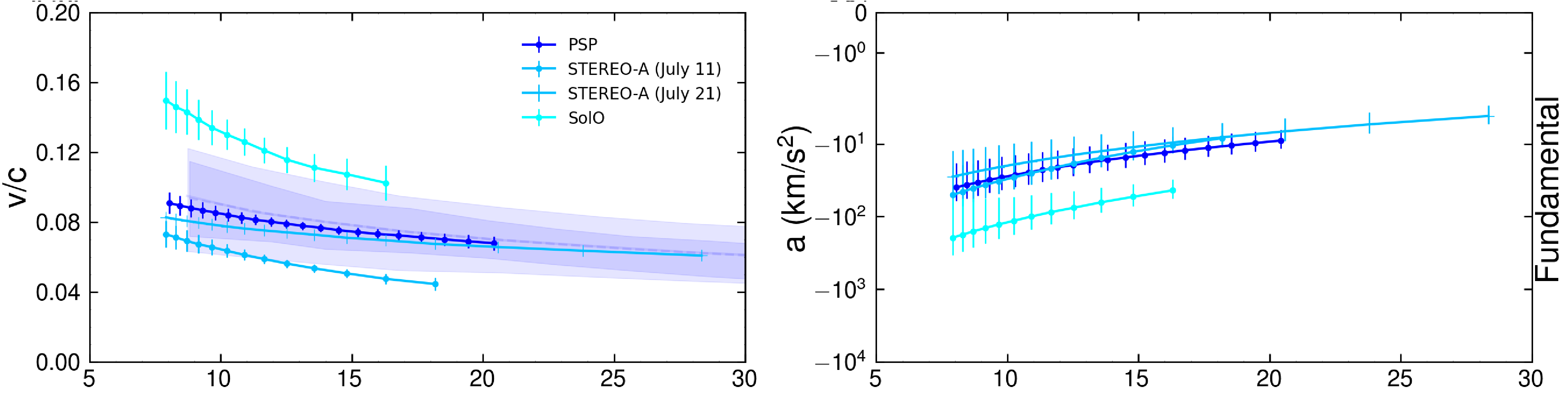}
\vskip 4pt
\includegraphics[width=0.9\linewidth]{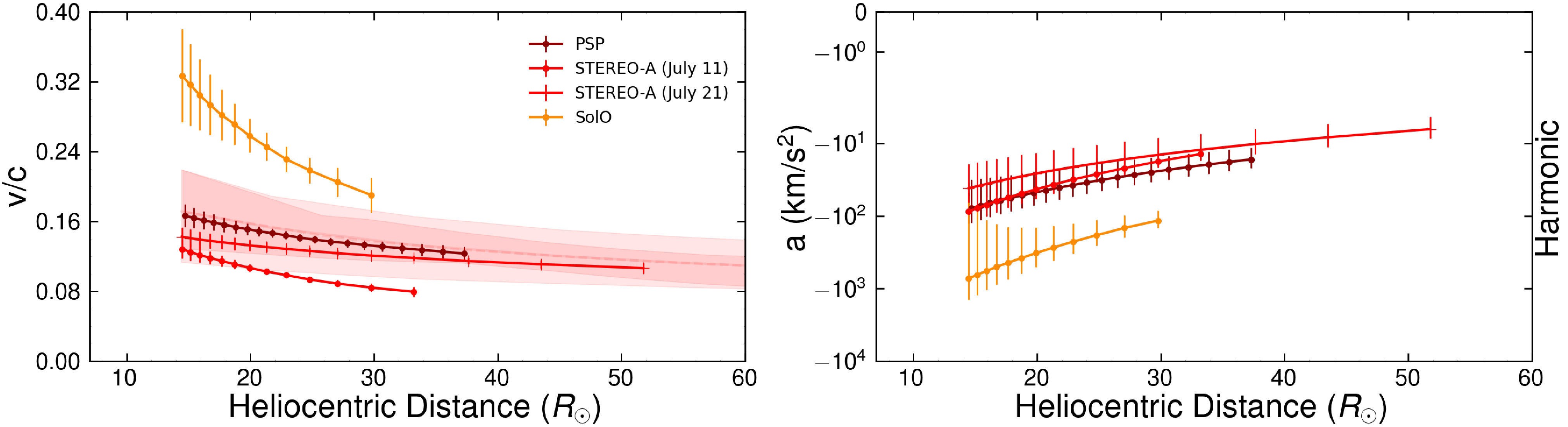}
   
\caption{Velocities deduced by PSP, STEREO-A and SolO data are
represented by lines different color shades, with darker to lighter
shades being associated to PSP, STEREO-A and SolO, respectively. Blue and red shades are associated to fundamental and harmonic emission, respectively. The top row displays fundamental emission, while the bottom row represents the
harmonic component. The events of July 11, 2020, and July 21, 2020, are
distinguished using dots and crosses as markers, respectively. Shaded regions represent the results from the \citet{2015A&A...580A.137K} analysis, with median values showcased by the transparent solid (STEREO-A) and dashed (STEREO-B) lines. 
}
\label{fig:psp_stereo}
\end{figure}

\section{Inhomogeneous Plasma and Electron Beam
Deceleration}\label{sec:theory}

The Langmuir waves driven by an electron beam are strongly affected by
density inhomogeneity. The solar corona and solar wind plasma is
inhomogeneous due to large scale density decrease with distance and due
to smaller scale density fluctuations. Inhomogeneities at both scales
will affect the Langmuir wave evolution via refraction and angular
scattering. While angular scattering of Langmuir waves changes the
direction of the wave-vector, refraction changes the wave-vector magnitude and hence the phase-speed. When the wavelength of a Langmuir
wave $\lambda$ is small compared to the size of the plasma inhomogeneity
\citep{1967PlPh....9..719V,1975PhFl...18..679C}, i.e.  $ \lambda \ll L$,
where
\begin{equation}\label{eq:L}
        L \equiv \left(\frac{1}{\omega_{pe}}\frac{\partial \omega_{pe}}{\partial x}\right)^{-1}= \left( \frac{\partial \ln \omega_{pe}}{\partial x}\right)^{-1},
\end{equation}
is the scale of ambient plasma density fluctuations, we can describe the
resonant interaction between the electron distribution function
$f(v,x,t)$ and the spectral energy density of Langmuir waves $W(v,x,t)$
through a system of kinetic equations
\citep[e.g.][]{1969JETP...30..131R,2001A&A...375..629K,2014A&A...572A.111R}
\begin{equation} \label{eq:kin1}
  \frac{\partial f}{\partial t}+v\frac{\partial f}{\partial x}=
  \frac{4\pi^2 e^2}{m^2}\frac{\partial
  }{\partial v}\frac{W}{v}\frac{\partial f}{\partial v}=\frac{\partial
  }{\partial v}D \frac{\partial f}{\partial v}
\end{equation}
\begin{equation}\label{eq:kin2}
  \frac{\partial W}{\partial t} + v_{gr} \frac{\partial W}{\partial x} - \frac{v^2}{L}\frac{\partial W}{\partial v}=\frac{\pi \omega _{pe}}{n_{e}}v^{2}W
  \frac{\partial f}{\partial v},
\end{equation}
where $\int Wdk=U$ and $\int f\text{d}v=n_b$ are the energy density of
Langmuir waves and the number density of the electron beam. Here, the
spontaneous terms are disregarded, as the beam-driven level of Langmuir
waves is significantly higher than the spontaneous or thermal level
\citep{1970JETP...31..396R}. The last two terms on the left-hand side of
Equation~(\ref{eq:kin2}) describe the propagation of Langmuir waves with
group velocity $v_{\text{gr}}<<v$ and refraction of wavenumber $k$. The
wave-number increases (phase speed decreases) when Langmuir waves
propagate into the region of decreasing plasma density
\citep{1967PlPh....9..719V,1969JETP...30..131R}. The right-hand side
terms of Equations~(\ref{eq:kin1},\ref{eq:kin2}) describe the dominant
resonant interaction $\omega_{p e}=k v$ between electrons with speed $v$
and plasma waves with wavenumber $k$. The $0^{\text {th }}$-order
solution is well known to be a plateau in the velocity space
(e.g.\citet{1967PlPh....9..719V,1970JETP...31..396R,1995PlPhR..21...89M,2000NewA....5...35M,2001A&A...375..629K,2024ApJ...976..233K}),
so the electron distribution function can be written as
$$
f(v, x, t)= \begin{cases}p(x, t), & 0<v<u(x, t) \\ 0, & v \geq u(x, t)\end{cases}
$$
and the spectral energy density of Langmuir waves as 
$$
W(v, x, t)= \begin{cases}W_0(v, x, t), & 0<v<u(x, t) \\ 0, & v \geq u(x, t),\end{cases}
$$
where $p\left(x, t\right)$ is the plateau height and $u(x, t)$ is the
maximum velocity of the electrons within the plateau. The electron
number density is the integral of the electron distribution function
over velocity
\begin{equation}\label{eq:p_norm}
  n(x,t)=\int\limits_{0}^{u(x,t)}p\left( x,t\right) \text{d}v=p\left( x,t\right) u(x,t)\,.
\end{equation}

Following \citep{1995PlPhR..21...89M,1999SoPh..184..353M}, one can
derive equations for $p(x,t)$, $u(x,t)$ and $W_0(v,x,t)$ using the
initial condition
\begin{equation}\label{eq:ft0}
  f(v,x,t=0)=n_b g(v) \exp(-x^2/d^2) \,,
\end{equation}
where $d$ and $n_b$ are the spatial size and the electron density of the
beam at $x=0$, and $g(v)=2v/v_0^2$ for $v<v_0$, these are 
\citep[see ][for details]{1995PlPhR..21...89M,2001CoPhC.138..222K,2024ApJ...976..233K}
\begin{align}
  &u(x,t) =  v_0 \,, \label{eq:solutionHD1}\\
  &p(x,t) = \frac{n_b}{v_0} \exp(-(x-v_0t/2)^2/d^2) \,,\label{eq:solutionHD2}\\
  &W_0(v,x,t)= \frac{m}{\omega_{pe}}v^4\left(1-\frac{v}{v_0}\right)p(x,t)\,,\label{eq:solutionHD3}
\end{align}
We note that in practice, the plateau is not extending to zero speed but
down to a few thermal speeds \citep[see discussion in
][]{2024ApJ...976..233K}. The solution for the Langmuir waves and
electrons allows to calculate the momentum density of the electron beam
$P_\text{b}$ and the momentum density of the Langmuir waves $P_\text{w}$
as the functions of $u(x,t)$. The momentum density of the electron beam
is the integral of the electron distribution function over velocity
\begin{equation}\label{eq:Pb}
  P_{\text{b}}\left( x,t\right) =\int\limits_{0}^{u(x,t)} m_e v p\left( x,t\right) \text{d}v =m_{e}n\left( x,t\right) \frac{u(x,t)}{2}\,,
\end{equation} 
and the momentum density of the Langmuir waves is the integral of the
spectral energy density of Langmuir waves multiplied by
$k=\omega_{pe}/v$ over velocity
\begin{equation}\label{eq:Pw}
  P_{\text{w}}\left( x,t\right) =\omega_{pe}\int\limits_{0}^{u(x,t)}\frac{W_0\left( v,x,t\right) }{v^{3}}\text{d}v =m_{e}\frac{n\left( x,t\right) }{u(x,t)}\int\limits_{0}^{u(x,t)}v\left( 1-\frac{v}{u(x,t)}\right) \text{d}v=m_{e}n\left( x,t\right) \frac{u(x,t)}{6}\,.
\end{equation}
The total momentum density of electrons and Langmuir waves is the sum of
Equations~(\ref{eq:Pb}) and (\ref{eq:Pw}):
\begin{equation}\label{eq:Ptot}
  P_{\text{tot}}\left( x,t\right) =P_{\text{b}}\left( x,t\right) +P_{\text{w}}\left( x,t\right) =m_{e}n\left( x,t\right) \frac{u(x,t)}{2}+m_{e}n\left( x,t\right) \frac{u(x,t)}{6}=m_{e}n\left( x,t\right) \frac{2 \,u(x,t)}{3}=4P_\text{w}\,.
\end{equation}
In the homogeneous plasma, the total momentum density
$P_{\text{tot}}=P_{\text{b}}+P_{\text{w}}$ is conserved
\citep{1999SoPh..184..353M}. In the inhomogeneous plasma, the situation
is more complicated. The numerical simulations by
\citet{2001A&A...375..629K,2013SoPh..285..217R,2014A&A...572A.111R} show
that, as the Langmuir waves propagate through a region of decreasing
background plasma density, they experience a negative shift in velocity
space towards smaller thermal speeds. This results in wave-absorption
(Landau dumping) by the Maxwellian component of the plasma and a
decrease in the total energy of the beam-wave structure. 
At the same time, the total momentum of the electron beam $P_\text{b}$ 
is a constant over the timescale of the shift in wave velocity, 
i.e. $\frac{\partial}{\partial t} P_{b} = 0$. 
The time evolution of electron
distribution and Langmuir waves can be seen in the Figure~1 by
\citet{2001A&A...375..629K}. 

Since the group speed of Langmuir waves $v_{gr}\ll v_{T_e}$, where
$v_{T_e}$ is the electron thermal velocity, we can ignore the spatial
motion of Langmuir waves and consider the effect of the refraction
assuming electrons have a plateau distribution, 
i.e. $\partial f/\partial v =0$. 
Then, we can simplify Equation~(\ref{eq:kin2}) as 
\begin{equation}\label{eq:kin4}
  \frac{\partial W}{\partial t}  - \frac{v^2}{L}\frac{\partial W}{\partial v}=0\,,
\end{equation}
where $L$ is the scale of ambient plasma density fluctuations defined by
Equation~(\ref{eq:L}). Multiplying Equation~(\ref{eq:kin4}) by
${\omega_{pe}}/{v^3}$ and integrating over velocity from 0 to $u$ we
obtain
\begin{equation}\label{eq:PW_inhom}
    \frac{\partial}{\partial t} \int_{0}^{u} \frac{\omega_{pe}}{v^3} W(v) dv - \int_{0}^{u} \frac{v^2 \omega_{pe}}{L v^3} \frac{\partial}{\partial v} W(v) dv =0\,,
\end{equation}
which reduces to
\begin{equation}\label{eq:PW_inhom2}
    \frac{\partial}{\partial t} P_W = \frac{m_e n u^2}{12 L} = \frac{u}{2L} P_W\,.
  \end{equation}
The equation shows that the momentum density of Langmuir waves, $P_W$, 
decreases for $L<0$, i.e. for decreasing density plasma. 
For $L>0$, i.e. increasing
density, the Langmuir waves increase velocity and can accelerate
electrons (see Figure~2 by \citet{2001A&A...375..629K}).
Noteworthy, the energy density of Langmuir waves remains constant, 
i.e.  multiplying Equation~(\ref{eq:kin4}) by
${\omega_{pe}}/{v^2}$ one finds $\frac{\partial}{\partial t} \int_{0}^{u} \frac{\omega_{pe}}{v^2} W(v) dv=0$.

The total momentum of the electron beam can be taken to be constant over
the timescale of the shift in wave velocity, i.e.
$\frac{\partial}{\partial t} P_{b} = 0$. Thus, we find for the total
momentum density of electrons and Langmuir waves changes as
\begin{equation}\label{eq:Ptot2}
  \frac{\partial}{\partial t} P_{\text{tot}} = \frac{\partial}{\partial t} P_{W} + \frac{\partial}{\partial t} P_{b} = \frac{u}{2L} P_W + 0 = \frac{u}{2L} \frac{P_{\text{tot}}}{4}\,.    
\end{equation}
i.e. the total momentum decreases due to the decrease in Langmuir wave
momentum density. Recalling the expression for the total momentum
density $P_{\text{tot}}=m_{e}n\frac{2 \,u}{3}$, for constant $n$ moving
with slowly changing speed $u$, we simplify Equation~(\ref{eq:Ptot2}) to
\begin{equation}\label{eq:Ptot3}
    \frac{\partial u}{\partial t}\simeq u\frac{\partial u}{\partial x}=\frac{u}{4 L}\,,
\end{equation}
which can be integrated with $L =
\left(\frac{1}{\omega_{pe}}\frac{\partial \omega_{pe}}{\partial
x}\right)^{-1}$ to give the solution for the velocity of the beam
plateau as a function of distance.
\begin{equation}\label{eq:u_x}
    \frac{u(x)}{u(x=x_0)} =\left(\frac{\omega_{pe}(x)}{\omega_{pe}(x=x_0)}\right)^{1/4} 
    =\left(\frac{f_{pe}(x)}{f_{pe}(x=x_0)}\right)^{1/4}\,.
\end{equation}

While simple and approximate, Equation (\ref{eq:u_x}) provides important
insight into the decrease of the electron speed due to decreasing
density. It shows that the decrease is faster for stronger decreasing
plasma. In solar wind with density $n(r)\propto r^{-2.3}$, the velocity
decreases  
as $u(r)\propto r^{-0.29}$ and acceleration changes as $a(r) \propto
r^{-1.58}$. Importantly, the model yields the proportionality $u \propto
\, f^{0.25}$, which is independent on the density model. Interestingly,
the acceleration $a(r)$ is rather steep function of $r$, $a(r) \propto
r^{-1.58}$, so the measured acceleration could differ by order of
magnitude for different frequencies and, in general, more sensitive to a
density model. The velocity decrease is probably unsurprisingly similar
to the numerical simulations
\citep{2001SoPh..202..131K,2013SoPh..285..217R,2023SoPh..298...52L} of
beam transport.

The comparison to velocity and acceleration estimations from type III
events analyzed in the previous section can be seen in
Figure~\ref{fig:11_07_speed_accel}, where Equation~\ref{eq:u_x} is over
plotted to the observational results as a dashed black curve. We note
the spread of the initial speed and accelerations, but similar $r$
dependency. While the beam deceleration model is in good agreement with
the observations, with predicted velocities and accelerations falling
within the margins of uncertainty for the observed $v \propto
f_{pe}^{0.32 \pm 0.12}$ and $a(r) \propto r^{-1.71 \pm 0.20}$, we note
the differences in values of speed/acceleration obtained by different
spacecrafts. The uncertainties comparable to the differences do not
allow firm conclusion, but it is tempting to suggest that the delay
$\delta t$ is influences by the scattering effects
\citep{2023ApJ...956..112K}, so the radio-waves are propagating slower
than $c$, which is implied by Equation \ref{eq:delta_t}.

\section{Summary}
We examine the drift rates of four type III radio bursts originating
from flares taking into account angular positions of the bursts. For the first time, simultaneous four-spacecraft observation allow inferred velocities and accelerations of type III emitters to be corrected for source-spacecraft angle. Exciter velocities are found to decrease with frequency as $u(f) \propto f^{0.32 \pm 0.12}$,
regardless of whether harmonic or fundamental emission is assumed; 
this is within the uncertainties to the case of a beam deceleration
propagating through background plasma of decreasing density that gives
speed $u(f) \propto f^{0.25}$. 

Assuming the density model in Equation~\ref{eq:density_r}, 
velocities are found to decrease with distance
for all events analyzed, with median $\beta_H \sim$ $\beta_F \sim$
-0.37 $\pm$ 0.15, values consistent with previous results published by
\citet{2015A&A...580A.137K}. Furthermore, exciter accelerations are
predicted to decrease faster with heliocentric distance as $a(r)
\propto r^{-1.58}$, in quite remarkable agreement with the observed
$a(r) \propto r^{-1.71 \pm 0.20}$. 
This result provides strong evidence for the interaction between beam-plasma
structure and density inhomogeneity being being the primary driver of
Type III solar radio burst exciter deceleration. 
It also lays a solid foundation for future work, 
which will likely involve
statistical analysis to reduce the uncertainness.

We also note that there are intriguing differences in the drift-rate of
the same type III bursts observed by different spacecrafts.  
The drift-rate analysis for the 11 July 2020 event using dynamic spectra
from PSP, STEREO-A and SolO spacecrafts show difference in velocities
recorded by different spacecrafts (Figure~\ref{fig:psp_stereo}).
Although the differences are only slightly exceed the uncertainties,
this discrepancy, not attributed to properties intrinsic to the exciter,
could be the result of radio waves scattering off density
inhomogeneities in the ambient plasma and affecting the type III burst
observed time characteristics \citep{2023ApJ...956..112K}. 
In other words, the measurement of the type III 
exciter deceleration using a single spacecraft could be a subject 
of noticeable error due to radio-wave scattering.

\begin{acknowledgments}
    The work was supported via the UKRI/STFC grant ST/Y001834/1. F.A.
    (studentship 2604774) and E.P.K. were supported via UKRI/STFC
    training grant ST/V506692/1. This research has made use of the
    Astrophysics Data System, funded by NASA under Cooperative Agreement
    80NSSC21M00561.
\end{acknowledgments}

\bibliography{refs}

\begin{thebibliography}{}
\expandafter\ifx\csname natexlab\endcsname\relax\def\natexlab#1{#1}\fi
\providecommand{\url}[1]{\href{#1}{#1}}
\providecommand{\dodoi}[1]{doi:~\href{http://doi.org/#1}{\nolinkurl{#1}}}
\providecommand{\doeprint}[1]{\href{http://ascl.net/#1}{\nolinkurl{http://ascl.net/#1}}}
\providecommand{\doarXiv}[1]{\href{https://arxiv.org/abs/#1}{\nolinkurl{https://arxiv.org/abs/#1}}}

\bibitem[{{Aschwanden} {et~al.}(1995){Aschwanden}, {Benz}, {Dennis}, \&
  {Schwartz}}]{1995ApJ...455..347A}
{Aschwanden}, M.~J., {Benz}, A.~O., {Dennis}, B.~R., \& {Schwartz}, R.~A. 1995,
  \apj, 455, 347, \dodoi{10.1086/176582}

\bibitem[{{Benz}(2017)}]{2017LRSP...14....2B}
{Benz}, A.~O. 2017, Living Reviews in Solar Physics, 14, 2,
  \dodoi{10.1007/s41116-016-0004-3}

\bibitem[{{Bonnin} {et~al.}(2008){Bonnin}, {Hoang}, \&
  {Maksimovic}}]{2008A&A...489..419B}
{Bonnin}, X., {Hoang}, S., \& {Maksimovic}, M. 2008, \aap, 489, 419,
  \dodoi{10.1051/0004-6361:200809777}

\bibitem[{{Bougeret} {et~al.}(1995){Bougeret}, {Kaiser}, {Kellogg}, {Manning},
  {Goetz}, {Monson}, {Monge}, {Friel}, {Meetre}, {Perche}, {Sitruk}, \&
  {Hoang}}]{1995SSRv...71..231B}
{Bougeret}, J.~L., {Kaiser}, M.~L., {Kellogg}, P.~J., {et~al.} 1995, \ssr, 71,
  231, \dodoi{10.1007/BF00751331}

\bibitem[{{Chen} {et~al.}(2023){Chen}, {Kontar}, {Chrysaphi}, {Zhang},
  {Krupar}, {Musset}, {Maksimovic}, {Jeffrey}, {Azzollini}, \&
  {Vecchio}}]{2023A&A...680A...1C}
{Chen}, X., {Kontar}, E.~P., {Chrysaphi}, N., {et~al.} 2023, \aap, 680, A1,
  \dodoi{10.1051/0004-6361/202347185}

\bibitem[{{Clarkson} {et~al.}(2025){Clarkson}, {Kontar}, {Chrysaphi}, {Emslie},
  {Jeffrey}, {Krupar}, \& {Vecchio}}]{2025NatSR..1511335C}
{Clarkson}, D.~L., {Kontar}, E.~P., {Chrysaphi}, N., {et~al.} 2025, Scientific
  Reports, 15, 11335, \dodoi{10.1038/s41598-025-95270-w}

\bibitem[{{Coste} {et~al.}(1975){Coste}, {Reinisch}, {Montes}, \&
  {Silevitch}}]{1975PhFl...18..679C}
{Coste}, J., {Reinisch}, G., {Montes}, C., \& {Silevitch}, M.~B. 1975, Physics
  of Fluids, 18, 679, \dodoi{10.1063/1.861192}

\bibitem[{{Dulk} {et~al.}(1987){Dulk}, {Goldman}, {Steinberg}, \&
  {Hoang}}]{1987A&A...173..366D}
{Dulk}, G.~A., {Goldman}, M.~V., {Steinberg}, J.~L., \& {Hoang}, S. 1987, \aap,
  173, 366

\bibitem[{{Fainberg} {et~al.}(1972){Fainberg}, {Evans}, \&
  {Stone}}]{1972Sci...178..743F}
{Fainberg}, J., {Evans}, L.~G., \& {Stone}, R.~G. 1972, Science, 178, 743,
  \dodoi{10.1126/science.178.4062.743}

\bibitem[{{Fainberg} \& {Stone}(1970)}]{1970SoPh...15..222F}
{Fainberg}, J., \& {Stone}, R.~G. 1970, \solphys, 15, 222,
  \dodoi{10.1007/BF00149487}

\bibitem[{{Fox} {et~al.}(2016){Fox}, {Velli}, {Bale}, {Decker}, {Driesman},
  {Howard}, {Kasper}, {Kinnison}, {Kusterer}, {Lario}, {Lockwood}, {McComas},
  {Raouafi}, \& {Szabo}}]{2016SSRv..204....7F}
{Fox}, N.~J., {Velli}, M.~C., {Bale}, S.~D., {et~al.} 2016, \ssr, 204, 7,
  \dodoi{10.1007/s11214-015-0211-6}

\bibitem[{{Ginzburg} \& {Zhelezniakov}(1958)}]{1958SvA.....2..653G}
{Ginzburg}, V.~L., \& {Zhelezniakov}, V.~V. 1958, Soviet~Ast., 2, 653

\bibitem[{{Hoang} {et~al.}(1997){Hoang}, {Poqu{\'e}russe}, \&
  {Bougeret}}]{1997SoPh..172..307H}
{Hoang}, S., {Poqu{\'e}russe}, M., \& {Bougeret}, J.~L. 1997, \solphys, 172,
  307, \dodoi{10.1023/A:1004956913131}

\bibitem[{{Holman} {et~al.}(2011){Holman}, {Aschwanden}, {Aurass}, {Battaglia},
  {Grigis}, {Kontar}, {Liu}, {Saint-Hilaire}, \&
  {Zharkova}}]{2011SSRv..159..107H}
{Holman}, G.~D., {Aschwanden}, M.~J., {Aurass}, H., {et~al.} 2011, \ssr, 159,
  107, \dodoi{10.1007/s11214-010-9680-9}

\bibitem[{{Hughes} \& {Harkness}(1963)}]{1963ApJ...138..239H}
{Hughes}, M.~P., \& {Harkness}, R.~L. 1963, \apj, 138, 239,
  \dodoi{10.1086/147630}

\bibitem[{{Kaiser}(2005)}]{2005AdSpR..36.1483K}
{Kaiser}, M.~L. 2005, Advances in Space Research, 36, 1483,
  \dodoi{10.1016/j.asr.2004.12.066}

\bibitem[{{Kontar}(2001{\natexlab{a}})}]{2001SoPh..202..131K}
{Kontar}, E.~P. 2001{\natexlab{a}}, Sol.~Phys., 202, 131

\bibitem[{{Kontar}(2001{\natexlab{b}})}]{2001A&A...375..629K}
---. 2001{\natexlab{b}}, \aap, 375, 629, \dodoi{10.1051/0004-6361:20010807}

\bibitem[{{Kontar}(2001{\natexlab{c}})}]{2001CoPhC.138..222K}
---. 2001{\natexlab{c}}, Computer Physics Communications, 138, 222,
  \dodoi{10.1016/S0010-4655(01)00214-4}

\bibitem[{{Kontar} {et~al.}(2024){Kontar}, {Azzollini}, \&
  {Lyubchyk}}]{2024ApJ...976..233K}
{Kontar}, E.~P., {Azzollini}, F., \& {Lyubchyk}, O. 2024, \apj, 976, 233,
  \dodoi{10.3847/1538-4357/ad8560}

\bibitem[{{Kontar} {et~al.}(2023){Kontar}, {Emslie}, {Clarkson}, {Chen},
  {Chrysaphi}, {Azzollini}, {Jeffrey}, \& {Gordovskyy}}]{2023ApJ...956..112K}
{Kontar}, E.~P., {Emslie}, A.~G., {Clarkson}, D.~L., {et~al.} 2023, \apj, 956,
  112, \dodoi{10.3847/1538-4357/acf6c1}

\bibitem[{{Krupar} {et~al.}(2015){Krupar}, {Kontar}, {Soucek}, {Santolik},
  {Maksimovic}, \& {Kruparova}}]{2015A&A...580A.137K}
{Krupar}, V., {Kontar}, E.~P., {Soucek}, J., {et~al.} 2015, \aap, 580, A137,
  \dodoi{10.1051/0004-6361/201425308}

\bibitem[{{Lecacheux} {et~al.}(1989){Lecacheux}, {Steinberg}, {Hoang}, \&
  {Dulk}}]{1989A&A...217..237L}
{Lecacheux}, A., {Steinberg}, J.~L., {Hoang}, S., \& {Dulk}, G.~A. 1989, \aap,
  217, 237

\bibitem[{{Ledenev}(2000)}]{2000SoPh..197..387L}
{Ledenev}, V.~G. 2000, \solphys, 197, 387, \dodoi{10.1023/A:1026516413624}

\bibitem[{{Lin}(1974)}]{1974SSRv...16..189L}
{Lin}, R.~P. 1974, \ssr, 16, 189, \dodoi{10.1007/BF00240886}

\bibitem[{{Lin}(1985)}]{1985SoPh..100..537L}
---. 1985, \solphys, 100, 537, \dodoi{10.1007/BF00158444}

\bibitem[{{Lorfing} \& {Reid}(2023)}]{2023SoPh..298...52L}
{Lorfing}, C.~Y., \& {Reid}, H. A.~S. 2023, \solphys, 298, 52,
  \dodoi{10.1007/s11207-023-02145-2}

\bibitem[{{Mann} {et~al.}(1999){Mann}, {Jansen}, {MacDowall}, {Kaiser}, \&
  {Stone}}]{1999A&A...348..614M}
{Mann}, G., {Jansen}, F., {MacDowall}, R.~J., {Kaiser}, M.~L., \& {Stone},
  R.~G. 1999, \aap, 348, 614

\bibitem[{{Mel'nik}(1995)}]{1995PlPhR..21...89M}
{Mel'nik}, V.~N. 1995, Plasma Physics Reports, 21, 89,
  \dodoi{10.48550/arXiv.1802.07806}

\bibitem[{{Melnik} {et~al.}(2011){Melnik}, {Konovalenko}, {Rucker}, {Boiko},
  {Dorovskyy}, {Abranin}, \& {Lecacheux}}]{2011SoPh..269..335M}
{Melnik}, V.~N., {Konovalenko}, A.~A., {Rucker}, H.~O., {et~al.} 2011,
  \solphys, 269, 335, \dodoi{10.1007/s11207-010-9703-4}

\bibitem[{{Mel'nik} \& {Kontar}(2000)}]{2000NewA....5...35M}
{Mel'nik}, V.~N., \& {Kontar}, E.~P. 2000, \na, 5, 35,
  \dodoi{10.1016/S1384-1076(00)00004-X}

\bibitem[{{Mel'nik} {et~al.}(1999){Mel'nik}, {Lapshin}, \&
  {Kontar}}]{1999SoPh..184..353M}
{Mel'nik}, V.~N., {Lapshin}, V., \& {Kontar}, E. 1999, \solphys, 184, 353,
  \dodoi{10.1023/A:1005191910544}

\bibitem[{{Melnik} {et~al.}(2015){Melnik}, {Brazhenko}, {Konovalenko},
  {Briand}, {Dorovskyy}, {Zarka}, {Frantsuzenko}, {Rucker}, {Rutkevych},
  {Panchenko}, {Denis}, {Zaqarashvili}, \&
  {Shergelashvili}}]{2015SoPh..290..193M}
{Melnik}, V.~N., {Brazhenko}, A.~I., {Konovalenko}, A.~A., {et~al.} 2015,
  \solphys, 290, 193, \dodoi{10.1007/s11207-014-0577-8}

\bibitem[{{M{\"u}ller} {et~al.}(2020){M{\"u}ller}, {St. Cyr}, {Zouganelis},
  {Gilbert}, {Marsden}, {Nieves-Chinchilla}, {Antonucci}, {Auch{\`e}re},
  {Berghmans}, {Horbury}, {Howard}, {Krucker}, {Maksimovic}, {Owen}, {Rochus},
  {Rodriguez-Pacheco}, {Romoli}, {Solanki}, {Bruno}, {Carlsson}, {Fludra},
  {Harra}, {Hassler}, {Livi}, {Louarn}, {Peter}, {Sch{\"u}hle}, {Teriaca}, {del
  Toro Iniesta}, {Wimmer-Schweingruber}, {Marsch}, {Velli}, {De Groof},
  {Walsh}, \& {Williams}}]{2020A&A...642A...1M}
{M{\"u}ller}, D., {St. Cyr}, O.~C., {Zouganelis}, I., {et~al.} 2020, \aap, 642,
  A1, \dodoi{10.1051/0004-6361/202038467}

\bibitem[{{Musset} {et~al.}(2021){Musset}, {Maksimovic}, {Kontar}, {Krupar},
  {Chrysaphi}, {Bonnin}, {Vecchio}, {Cecconi}, {Zaslavsky}, {Issautier},
  {Bale}, \& {Pulupa}}]{2021A&A...656A..34M}
{Musset}, S., {Maksimovic}, M., {Kontar}, E., {et~al.} 2021, \aap, 656, A34,
  \dodoi{10.1051/0004-6361/202140998}

\bibitem[{{Parker}(1960)}]{1960ApJ...132..821P}
{Parker}, E.~N. 1960, \apj, 132, 821, \dodoi{10.1086/146985}

\bibitem[{{Poqu{\'e}russe} {et~al.}(1996){Poqu{\'e}russe}, {Hoang}, {Bougeret},
  \& {Moncuquet}}]{1996AIPC..382...62P}
{Poqu{\'e}russe}, M., {Hoang}, S., {Bougeret}, J.~L., \& {Moncuquet}, M. 1996,
  in American Institute of Physics Conference Series, Vol. 382, Proceedings of
  the eigth International solar wind Conference: Solar wind eight, ed.
  D.~{Winterhalter}, J.~T. {Gosling}, S.~R. {Habbal}, W.~S. {Kurth}, \&
  M.~{Neugebauer} (AIP), 62--65, \dodoi{10.1063/1.51360}

\bibitem[{{Press} {et~al.}(1986){Press}, {Flannery}, \&
  {Teukolsky}}]{1986nras.book.....P}
{Press}, W.~H., {Flannery}, B.~P., \& {Teukolsky}, S.~A. 1986, {Numerical
  recipes. The art of scientific computing} (Cambridge: University Press)

\bibitem[{{Ratcliffe} {et~al.}(2014){Ratcliffe}, {Kontar}, \&
  {Reid}}]{2014A&A...572A.111R}
{Ratcliffe}, H., {Kontar}, E.~P., \& {Reid}, H.~A.~S. 2014, \aap, 572, A111,
  \dodoi{10.1051/0004-6361/201423731}

\bibitem[{{Reid} \& {Kontar}(2013)}]{2013SoPh..285..217R}
{Reid}, H. A.~S., \& {Kontar}, E.~P. 2013, \solphys, 285, 217,
  \dodoi{10.1007/s11207-012-0013-x}

\bibitem[{{Ryutov}(1969)}]{1969JETP...30..131R}
{Ryutov}, D.~D. 1969, Soviet Journal of Experimental and Theoretical Physics,
  30, 131

\bibitem[{{Ryutov} \& {Sagdeev}(1970)}]{1970JETP...31..396R}
{Ryutov}, D.~D., \& {Sagdeev}, R.~Z. 1970, Soviet Journal of Experimental and
  Theoretical Physics, 31, 396

\bibitem[{{Thompson}(2006)}]{2006A&A...449..791T}
{Thompson}, W.~T. 2006, \aap, 449, 791, \dodoi{10.1051/0004-6361:20054262}

\bibitem[{{Vedenov} {et~al.}(1967){Vedenov}, {Gordeev}, \&
  {Rudakov}}]{1967PlPh....9..719V}
{Vedenov}, A.~A., {Gordeev}, A.~V., \& {Rudakov}, L.~I. 1967, Plasma Physics,
  9, 719, \dodoi{10.1088/0032-1028/9/6/305}

\bibitem[{{Wild}(1950)}]{1950AuSRA...3..541W}
{Wild}, J.~P. 1950, Australian Journal of Scientific Research A Physical
  Sciences, 3, 541, \dodoi{10.1071/CH9500541}

\end{thebibliography}

\appendix

\section{21 July 2020 Event}\label{app:July21}

\begin{figure}[h]
\centering
   \includegraphics[width=0.9\linewidth]{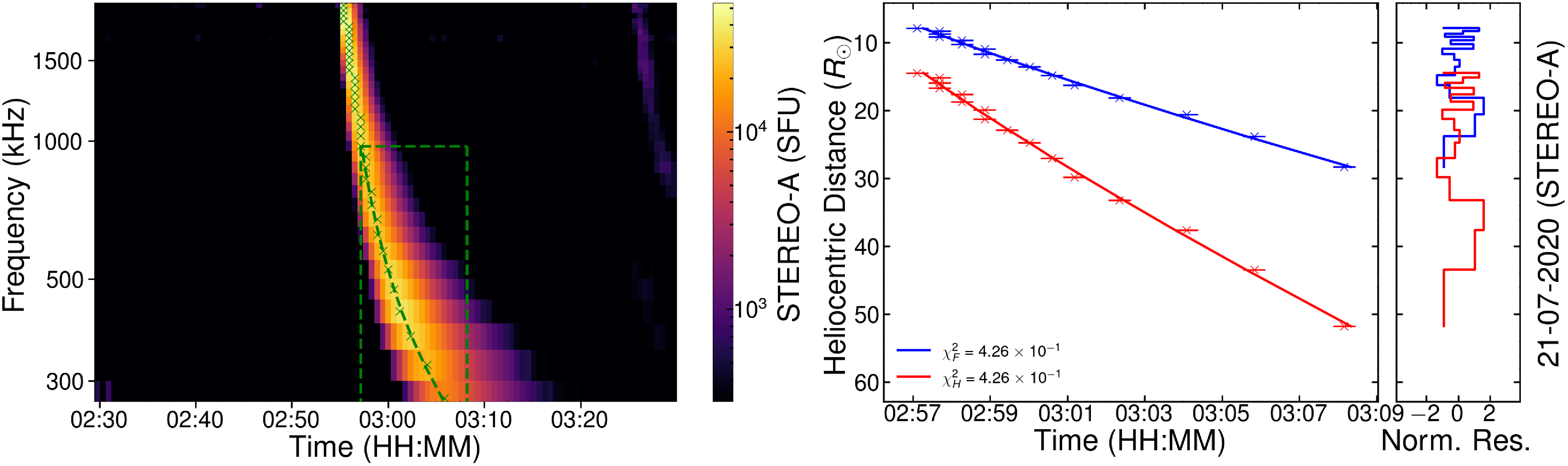}
   \vskip 4pt
  \includegraphics[width=0.9\linewidth]{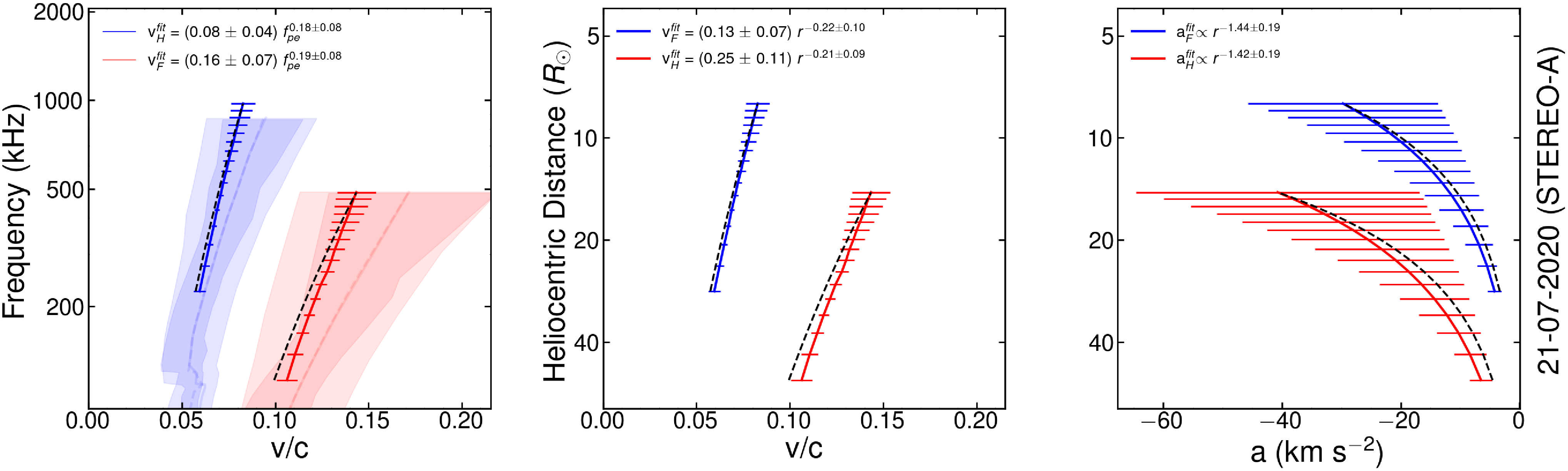}
   \caption{\label{fig:21_07_dynamic_spectra} The same as
   ~Figure~\ref{fig:11_07_dynamic_spectra} (top) and ~Figure~\ref{fig:11_07_speed_accel} (bottom), but for the 21 July 2020
   event observed by the STEREO-A spacecraft.}
\end{figure}

A type III event is observed by the four spacecraft 
on the 21 July 2020 03:00 UT. Longitude of maximum directivity is estimated to be $\sim 150^\circ$ in the HEE coordinate system, in agreement with \cite{2021A&A...656A..34M}. Regrettably, in this time-frame only STEREO-A is capturing spectral data with high enough time resolution to aid in our investigation. Results are shown in Figure~\ref{fig:21_07_dynamic_spectra}.

\begin{figure}
    \centering
    \includegraphics[width=0.9\linewidth]{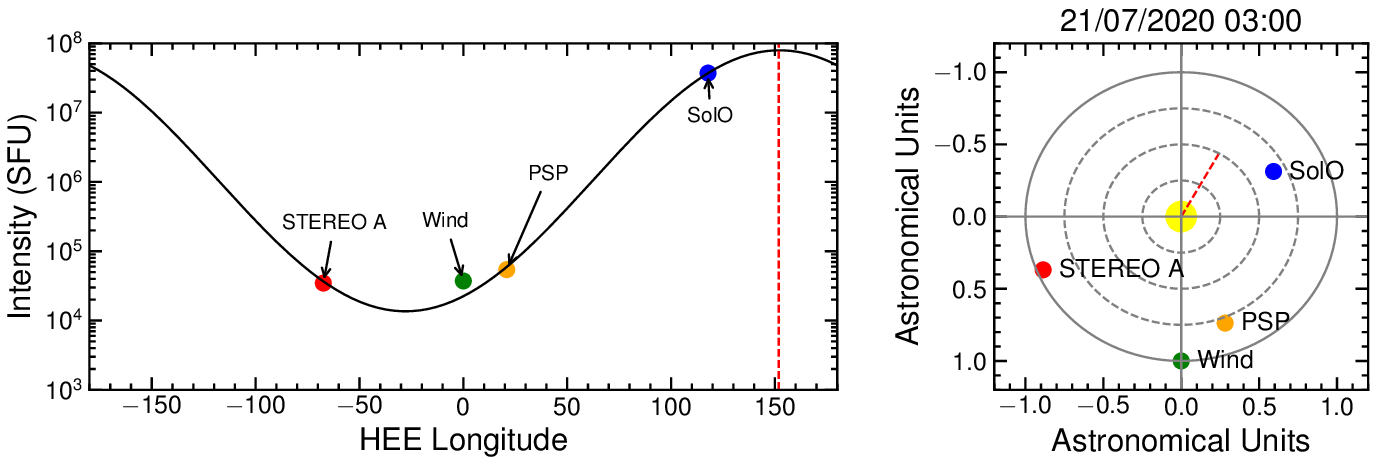}
    \caption{The same as
   ~Figure~\ref{fig:sc_position_11July} but for the 21 July 2020
   event observed by the STEREO-A spacecraft.}
    \label{fig:sc_position_July21}
\end{figure}

\end{document}